\documentclass[trackchanges]{aastex701}

\begin{document}

\title{Highly filamentary H{\,\small I} gas in the circumgalactic medium and intragalactic medium around NGC 4631}

\author{C. Zhang}
\affiliation{Department of Physics, Taiyuan Normal University, Jinzhong 030619, China}
\affiliation{Institute of Computational and Applied Physics, Taiyuan Normal University, Jinzhong 030619, China}
\email[show]{zhangchao920610@126.com}  

\author{Liu Tie} 
\affiliation{State Key Laboratory of Radio Astronomy and Technology, Shanghai Astronomical Observatory, Chinese Academy of Sciences, \\
80 Nandan Road, Shanghai 200030, People's Republic of China}
\email[show]{liutie@shao.ac.cn}

\author{Xiaofeng Mai}
\affiliation{State Key Laboratory of Radio Astronomy and Technology, Shanghai Astronomical Observatory, Chinese Academy of Sciences, \\
80 Nandan Road, Shanghai 200030, People's Republic of China}
\email[show]{maixf@shao.ac.cn}

\author{Jing Wang}
\affiliation{Kavli Institute for Astronomy and Astrophysics, Peking University, Beijing 100871, People's Republic of China}
\email[show]{jwang\_astro@pku.edu.cn}

\begin{abstract}
Neutral hydrogen (H{\,\small I}) in the circumgalactic medium (CGM) and intergalactic medium (IGM) traces baryon cycling and galaxy evolution, yet fine filamentary substructures in diffuse CGM/IGM H{\,\small I} remain poorly constrained observationally. We combine FAST FEASTS single-dish and VLA THINGS interferometric H{\,\small I} data of the NGC 4631 group to report the first robust detection of kpc-scale filamentary structures in the CGM/IGM, with widths of $0.5$-$3.3\ \mathrm{kpc}$ and lengths of $6.1$-$49.8\ \mathrm{kpc}$. These features confirm that low-density CGM/IGM gas hosts velocity coherent substructure. From position-velocity kinematic analysis, we identify three filament classes (U-shaped, linear, and wavy), implying diverse formation mechanisms. Our results establish a structural bridge connecting pc-scale interstellar filaments, kpc-scale CGM/IGM filaments, and Mpc-scale cosmic-web filaments, providing key observational support for multiscale gaseous coupling in cosmic ecosystem.
\end{abstract}

\keywords{\uat{Galaxies}{573} --- \uat{Intergalactic medium}{813} --- \uat{Circumgalactic medium}{1879}}

\section{Introduction} 

The circumgalactic medium (CGM) and intergalactic medium (IGM) constitute the critical interface between individual galaxies and the large-scale cosmic web, governing baryon cycling and galaxy evolution \citep{Keres10.1111/j.1365-2966.2005.09451.x,Putman2012ARA&A..50..491P,Tumlinson2011ApJ...733..111T}. As the dominant tracer of cool, dense baryonic gas, neutral hydrogen (H{\small I}) sustains ongoing star formation and shapes galaxy evolution through gas accretion. The spatial distribution and physical properties of H{\small I} in the CGM/IGM therefore offer fundamental insights into galaxy–cosmic web interactions \citep{Saintonge2017ApJS..233...22S,Borthakur2016ApJ...833..259B,Lan2018ApJ...866...36L,Blumenthal2018MNRAS.479.3952B,Wang2020ApJ...890...63W,Yu2022ApJ...930...85Y}.

While Mpc-scale filaments in the cosmic web have been well characterized by simulations and large surveys (e.g., \citealt{CHENdoi:10.1142/S2010194512006459,Li_2023,Meng2026ApJ..1002..210M}), high-resolution spectroscopic observations of fine filamentary structures within the CGM and IGM remain scarce. Within the Milky Way, filaments are ubiquitous in molecular clouds at parsec scale or even below sub-parsec scale \citep{2010A&A...518L.102A,2014prpl.conf...27A,2022MNRAS.514.6038Z,Hacar2023ASPC..534..153H,2026arXiv260211617Z,2026arXiv260404501Z}. Systematic observations of the ``giant" filaments in the interstellar medium (ISM) have also been conducted \citep{Hacar2023ASPC..534..153H}, revealing lengths of tens to hundreds of parsecs \citep{Ragan2014A&A...568A..73R,Zucker2015ApJ...815...23Z,Zhang2019A&A...622A..52Z,Colombo2021A&A...655L...2C,2025NatAs...9.1366L}, with the longest potentially extending up to $\sim$2 kpc \citep{Veena2021ApJ...921L..42V}. In contrast, high-resolution observations of cold gas filaments in the extragalactic CGM and IGM are exceedingly rare. For instance, \cite{Lin2025ApJ...995...12L} detected filamentary gas inflow in galaxy J0910b, and \cite{Melendez2015ApJ...804...46M} observed a $\sim$6 kpc outflow-driven filament in NGC 4631. Other filaments may arise from different physical mechanisms. Tidal interactions can generate large-scale H{\small I} filaments, as demonstrated by the $\sim$160 kpc tidal structure detected in the compact group HCG 16 \citep{OSullivan2014ApJ...793...74O,Jones2019}. Compressive turbulence, meanwhile, has been shown in numerical simulations to produce large-scale filamentary structures in multiphase gas \citep{Mohapatra2022MNRAS.514.3139M,2025NatAs...9.1366L}. 

Although various mechanisms (e.g., tidal force, turbulence, outflow, inflow) are physically distinct, they all yield filamentary morphologies, tracing different aspects of the multi-scale baryonic cycle. However, whether kpc-scale H{\small I} filaments are ubiquitously present in the CGM and IGM remains an open question. A major obstacle to studying gas structures in the CGM and IGM is the intrinsically low density of diffuse gas, which poses a formidable detection challenge for conventional telescopes. The Five-hundred-meter Aperture Spherical radio Telescope (FAST), with its unparalleled sensitivity to faint H{\small I} emission, has enabled global CGM/IGM surveys and statistical studies of their bulk properties \citep{Jiang2019SCPMA..6259502J, Wang2023ApJ...944..102W}.

The nearby edge-on barred spiral galaxy NGC 4631 (the Whale Galaxy; $\sim$7.5 Mpc; \citealt{Wang2023ApJ...944..102W}) provides an optimal target for studying gas structures within the IGM and CGM. The galaxy hosts vigorous star formation and is surrounded by an extended diffuse H{\small I} envelope exceeding 120 kpc in extent, plausibly condensed from the hot IGM. Complementarily, its circumgalactic environment already exhibits confirmed filamentary features in infrared tracers \citep{Melendez2015ApJ...804...46M}, rendering NGC 4631 a unique laboratory to disentangle whether diffuse CGM/IGM H{\small I} is homogeneous or organized into discrete filaments.

In this work, we present high-resolution and high-sensitivity H{\small I} observations of NGC 4631. We aim to probe fine-scale substructures in its surrounding CGM and IGM, and to distinguish whether the low-density gaseous halo is featureless or dominated by filamentary networks. This study provides the first systematic characterization of CGM/IGM H{\small I} filaments around NGC 4631, fills the observational gap in cool gas structures at the galaxy–cosmic web interface, and supplies empirical constraints for multiscale gas formation theories and galaxy–web interaction models.

\section{Data} 

This study utilizes the high-quality H{\small I} observations of the NGC 4631 galaxy group published by \citet{Wang2023ApJ...944..102W}. These data were obtained from single-dish observations with the Five-hundred-meter Aperture Spherical radio Telescope (FAST) and interferometric observations with the WSRT, targeting the tidally interacting NGC 4631 and its surrounding group. The FAST data are part of the FEASTS survey (project ID: PT2021\_0071), with publicly available products, processing documentation, and supplementary materials for the LVgal project. The WSRT data are part of the HALOGAS survey \citep{Heald2011A&A...526A.118H}.

In \citet{Wang2023ApJ...944..102W}, the FAST data were processed through a standard pipeline including RFI flagging and excision, bandpass and flux calibration, imaging gridding, and baseline flattening. The FAST data were projected onto the same celestial coordinates and WCS system as the WSRT HALOGAS data to ensure strict spatial alignment. The combined dataset enables both high-resolution imaging of compact H{\small I} structures and high-sensitivity detection of extended diffuse gas, making it well suited for studying CGM and IGM gas structures. In this work, we use the combined FAST and WSRT data, which has an angular resolution of $\sim45\arcsec\times39\arcsec$, a pixel size of 4$\arcsec$ and a velocity resolution of $\sim$4 km s$^{-1}$. More details on data reduction were presented in \citet{Wang2023ApJ...944..102W}. 

\section{Results}

\subsection{Identification of Filamentary Structures}

We employ the Subspace Constrained Mean Shift (SCMS) algorithm to extract filamentary structures from the HI data of the NGC 4631 galaxy group. The core workflow is summarized as follows. First, deep learning-based Gaussian fitting is applied to the spectral line data for denoising the position-position-velocity (PPV) cube (details in Mai et al., in preparation). This step effectively separates genuine astronomical signals from noise, especially in the circumgalactic medium (CGM) and intergalactic medium (IGM) regions featured by faint emission and heavy noise contamination, thus significantly enhancing the detectability of weak filaments.

Subsequently, ridge points are extracted from the denoised data using the SCMS algorithm. By tracing local maxima in the data density gradient field, SCMS precisely locates the central ridges of filamentary structures, and maintains high positional accuracy in both spatial and velocity dimensions, even in regions with complex gas distributions, velocity gradients, and overlapping gas components. Finally, the Minimum Spanning Tree (MST) algorithm is adopted to identify the longest connected paths among ridge points and reconstruct complete filamentary morphologies. The MST method efficiently eliminates isolated noise spikes and discontinuous ridge fragments, guaranteeing the physical continuity and integrity of the extracted filaments. Detailed algorithm implementations will be reported in Zhang et al. (in preparation).

\begin{figure*}[bthp!]
\includegraphics[width=1.0\textwidth]{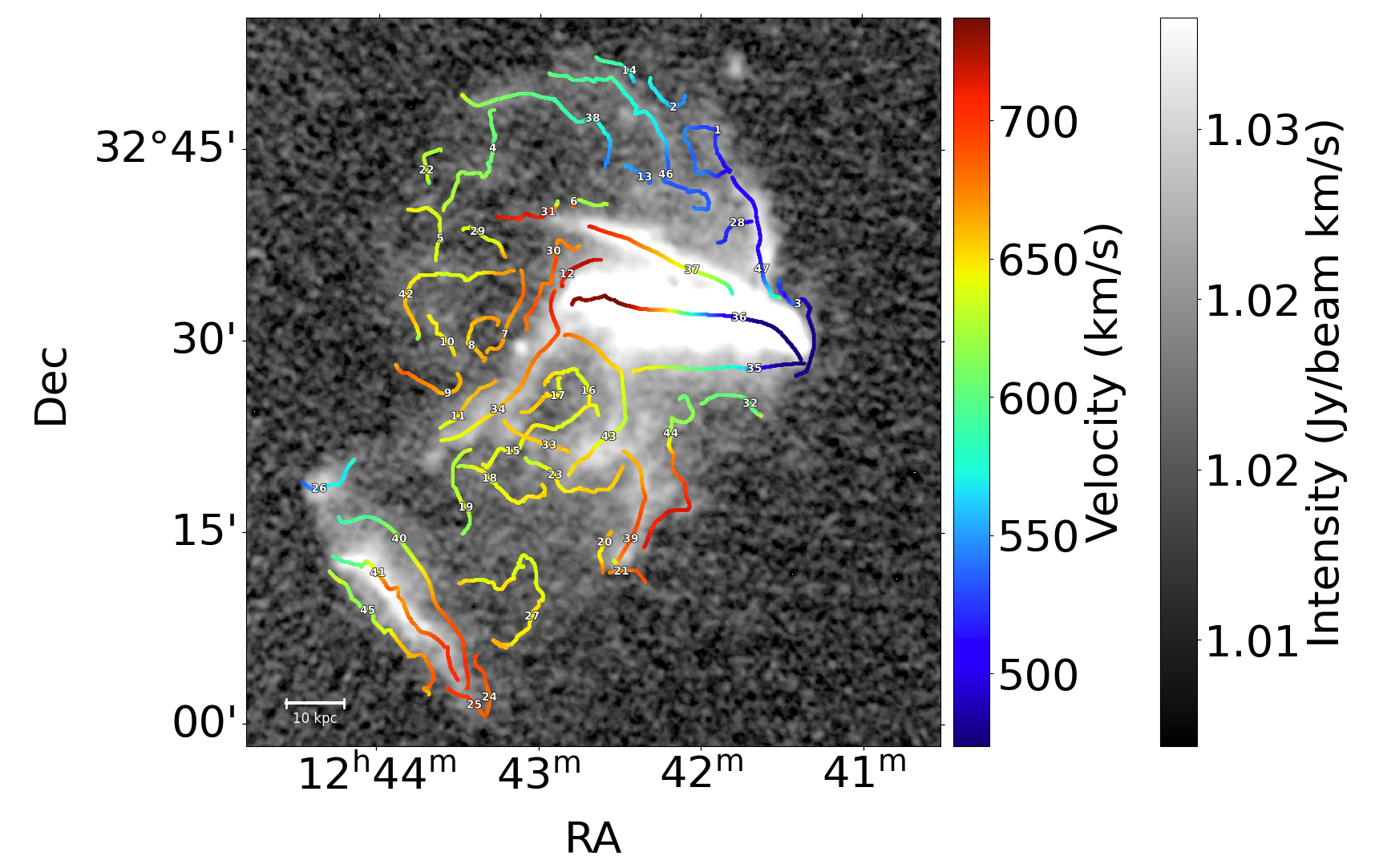}
\caption{The gray background shows the intensity map of the NGC 4631 galaxy group, and colored lines denote the extracted filaments. The number indicated on each filament represents its serial number.}
\label{fig:skeleton}
\end{figure*}

A length threshold of 4 times the beam size (\(1\ \mathrm{beam} = 10\ \mathrm{pixels}\)) is adopted for filament identification to retain statistically significant structures. This criterion ensures the detected filaments are spatially well resolved and excludes short artificial structures dominated by noise. In total, 47 filamentary structures are extracted. Figure~\ref{fig:skeleton} presents the final results, where the gray background represents the HI intensity map of the NGC 4631 galaxy group, and colored lines indicate the identified filaments. Different colors correspond to distinct velocity components, and continuous color gradients trace the systematic gas motion along the filaments.

Two prominent filaments extend across the galactic disks of NGC 4631 and NGC 4656, delineating the disk skeletons. Beyond the high-intensity disk regions, elongated filaments are also clearly detected in the surrounding CGM and IGM. These CGM/IGM filaments exhibit substantially lower HI column densities compared to those within the galactic disks.

\subsection{Physical properties of filamentary structures}
\subsubsection{Widths and lengths}
To validate the reliability of the extracted filaments and eliminate spurious structures generated by the algorithm, we analyze the transverse density profile of each filament. This verification is critical, as ridge-detection algorithms may produce artificial features under noisy conditions or smooth intensity gradients. A Gaussian or quasi-Gaussian profile signifies a well-defined central density peak and clear boundary, corresponding to a physically genuine filament. Conversely, profiles exhibiting random fluctuations or flat distributions are likely dominated by noise or represent algorithmic artifacts.

We employ the \textit{radfil}\footnote{\url{https://github.com/catherinezucker/radfil}} package to derive transverse density profiles for all identified filaments. The filaments analyzed in this study are three-dimensional structures containing velocity information. To mitigate line-of-sight projection effect, we compute the moment-0 (integrated intensity) map within the velocity range specific to each individual filament. \textit{radfil} extracts transverse intensity profiles at each position along the filament skeleton. We subsequently apply asymmetric reweighted penalized least squares (ARPLS) smoothing for baseline fitting of the intensity profiles. After that, we compute the median of the baseline-subtracted profiles across the filament skeleton to generate a representative transverse density profile. The median profile is then fitted with a Gaussian function, from which the characteristic width of each filament is determined as the full width at half maximum (FWHM). A representative example of filament width fitting is shown in Figure~\ref{fig:density_profile}; analogous fitting results for all filaments are provided in the supplementary materials.

\begin{figure*}[bthp!]
\includegraphics[width=1.0\textwidth]{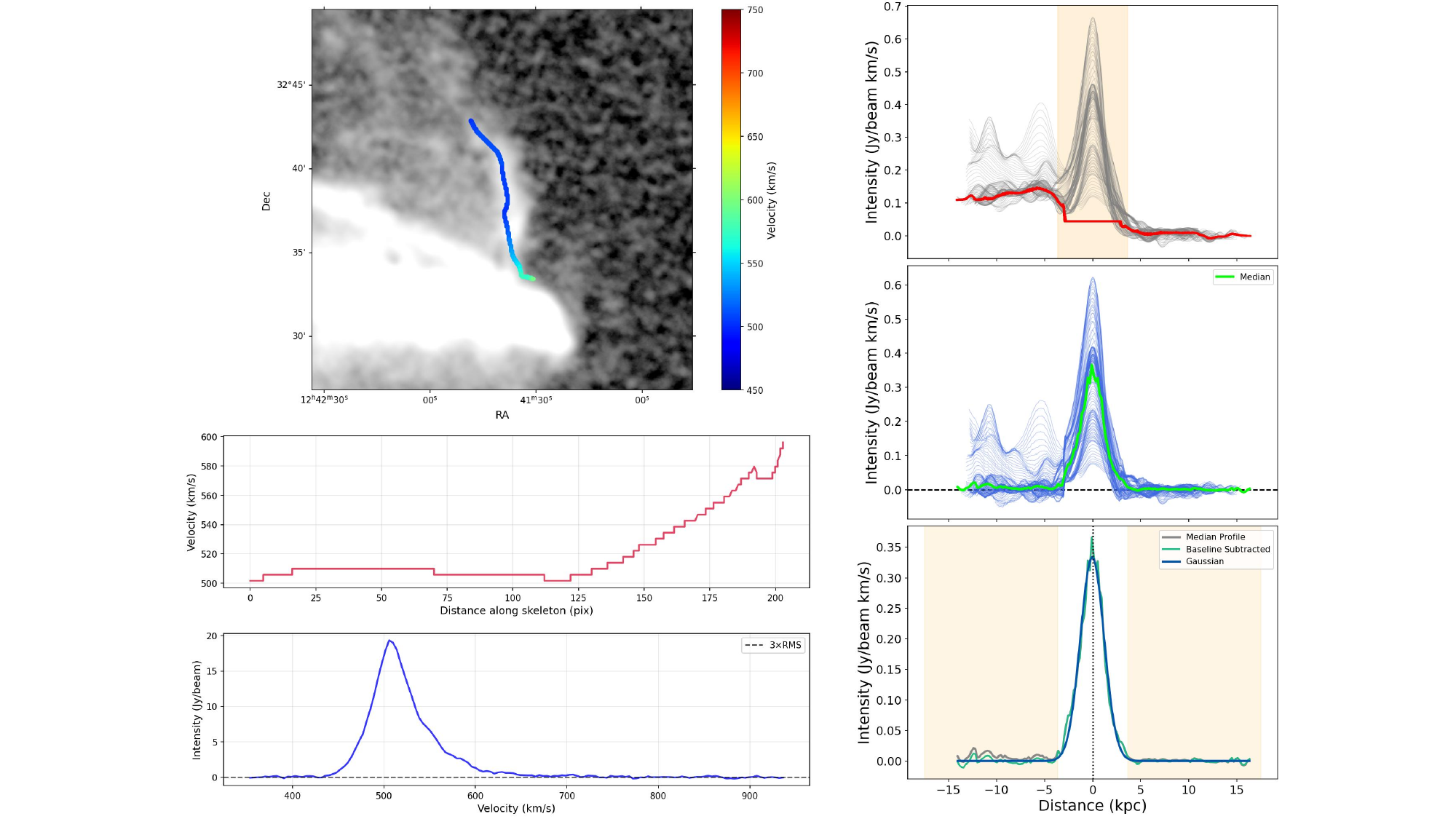}
\caption{Filament 47 as an example. Panel (a) shows the spatial distribution of the identified filaments. The gray background represents the intensity map, and the colored lines indicate the filamentary structures, with different colors corresponding to distinct velocity components. Panel (b) presents the velocity profile along the filament. Panel (c) displays the average spectral profiles averaged over a region spanning one beam width on either side of the filament skeleton; the black dashed line marks the 3$\sigma$ noise level. In Panel (d), the gray line shows the transverse density profile before baseline subtraction, while the red line represents the baseline fitted using the arpls algorithm. Panel (e) shows the transverse density profile after baseline subtraction (blue line) together with the extracted median profile (green line). Panel (f) presents the median transverse density profile (gray line) and its Gaussian fit (blue line).}
\label{fig:density_profile}
\end{figure*}

Seven filaments are found to be nearly parallel to the galactic disk and lie close to the disk plane. Owing to the intense HI emission from the disk, the transverse profiles of these filaments can only be reliably extracted on the side facing away from the disk; the disk-facing side suffers severe contamination, resulting in only half of a valid profile. For these filaments, we still perform Gaussian fitting to derive their widths, but label them as disk-affected and exclude them from the subsequent statistical analysis. Furthermore, the two filaments located within the galactic disks of NGC 4631 and NGC 4656 are also excluded from the CGM/IGM statistical analysis, as the dynamical environment within galactic disks is fundamentally distinct from that of the diffuse CGM/IGM gas. Images of these two disk filaments are provided in the Appendix.

\begin{figure*}[bthp!]
\includegraphics[width=1.0\textwidth]{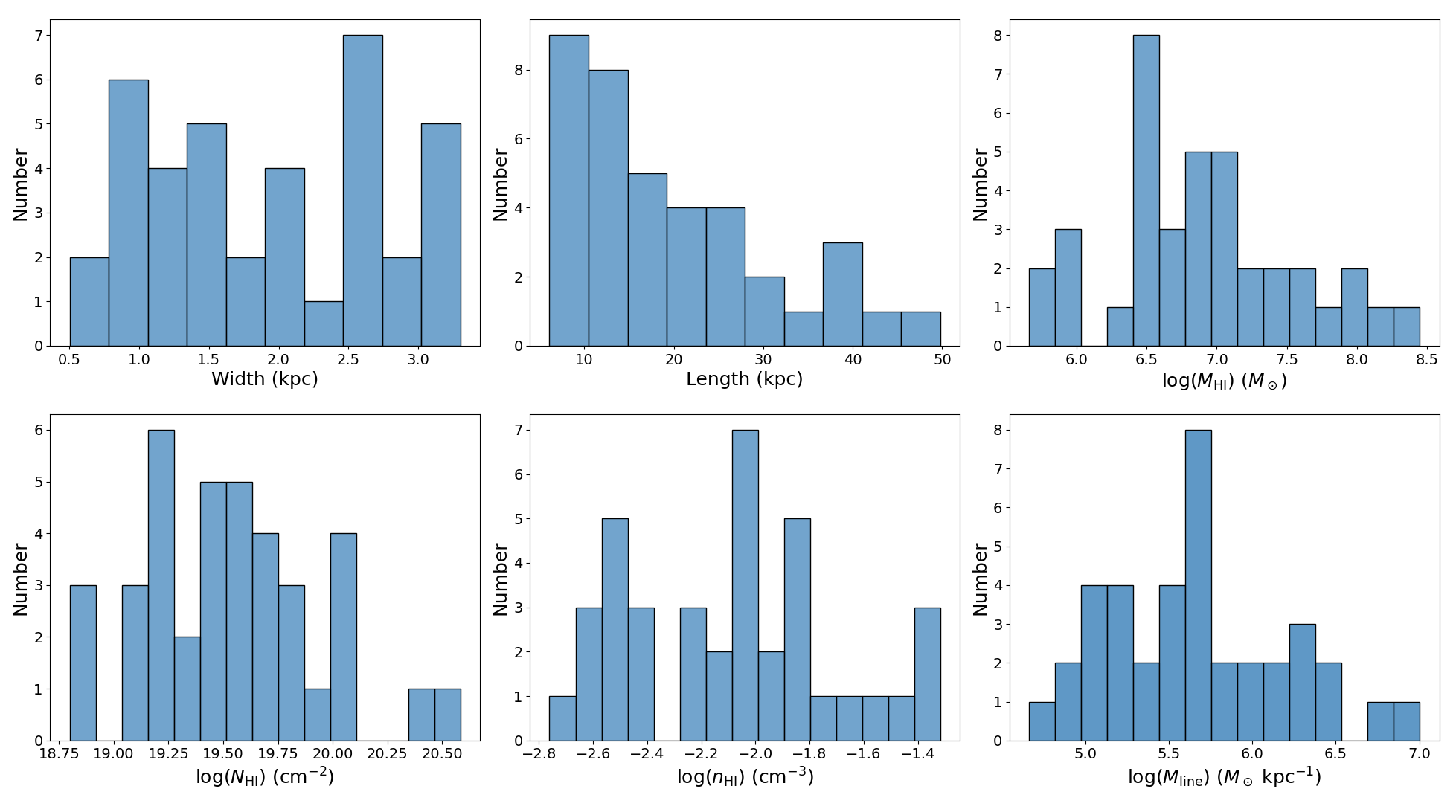}
\caption{Distributions of the physical properties of identified H{\,\small I} filaments in NGC 4631. Panel (a) shows the distribution of filament width, panel (b) shows the distribution of filament length, and panel (c) shows the logarithmic distribution of H{\,\small I} mass. Panel (d) presents the logarithmic distribution of H{\,\small I} median column density, panel (e) shows the logarithmic distribution of H{\,\small I} volume density, and panel (f) displays the logarithmic distribution of H{\,\small I} linear mass density.}
\label{fig:all_prameters}
\end{figure*}

The measured filament widths are summarized in Table~\ref{tab:filament_params}. Filaments with half-profiles due to disk contamination are marked with an asterisk (*), and the two disk filaments are denoted by double asterisks (**). After deconvolving the beam size, the intrinsic (deconvolved) widths of the CGM/IGM filaments span from 0.5 kpc to 3.3 kpc, with a median of $\sim$1.8 kpc and a mean of $\sim$1.9 kpc. The distribution of filament widths is displayed in Figure~\ref{fig:all_prameters}.

Filament lengths range from 6.1 kpc to 49.8 kpc, with a median of 17.4 kpc and a mean of 19.8 kpc. The fractions of filaments with aspect ratios (length/width) exceeding 3, 5, and 10 are 97.4\%, 86.8\%, and 39.5\%, respectively. These morphological properties confirm that the extracted CGM/IGM structures correspond to bona fide filaments, rather than randomly connected noise clumps or projection artifacts. Our results directly demonstrate that the low-density gas in the CGM and IGM possesses well-defined filamentary structures with spatial extents up to tens of kpc.

\subsubsection{Densities and masses}

Using the measured lengths and widths, we derive the average column density, volume density, mass, and linear mass (mass per unit length) for each filament. We adopt a cylindrical geometry for all calculations. The cylinder diameter is set to the Gaussian full width at half maximum (FWHM) derived from fitting the transverse HI column density profile, while its length traces the filamentary skeleton. Volume density is computed by dividing the column density by the cylinder diameter, under the assumption that the line-of-sight depth matches the projected transverse width. The total gas mass is calculated via integration of the column density field, and the linear mass is then defined as the total mass normalized by the filament length.

Statistical properties of the CGM/IGM filaments are summarized as follows. The average HI column density $\log(N_{\rm HI}/{\rm cm}^{-2})$ spans 18.8--20.6, with a median of 19.5 and mean of 19.6. The volume density $\log(n_{\rm HI}/{\rm cm}^{-3})$ ranges from $-2.8$ to $-1.3$, with median and mean both equal to $-2.1$. The mass $\log(M_{\rm HI}/M_\odot)$ covers 5.7--8.4, with a median of 6.9 and mean of 7.0. The linear mass $\log(M_{\rm line}/M_\odot\,{\rm kpc}^{-1})$ falls in the range 4.7--7.0, with a median of 5.6 and mean of 5.7. Adopting a typical temperature of $10^4$ K for CGM/IGM gas, the critical line mass is $m_{\rm crit}\sim1.6\times10^7\ M_\odot/\mathrm{kpc}$, implying that the majority of detected filaments are subcritical. Detailed values are listed in Table~\ref{tab:filament_params}, and parameter distributions are shown in Figure~\ref{fig:all_prameters}.

The CGM/IGM filaments differ distinctly from NGC 4631 disk filaments. Typical disk HI column densities reach $\sim10^{21}\ \mathrm{cm^{-2}}$, while CGM/IGM values are lower by 0.5 to 2 orders of magnitude, consistent with a more diffuse gas phase. The total mass of all CGM/IGM filaments is $\sim9.5\times 10^{8}\ M_\odot$. Relative to the total HI mass of NGC 4631 ($1.20\times 10^{10}\ M_\odot$; \citet{Wang2023ApJ...944..102W}), these filaments contribute $\sim7.9\%$ of the galactic HI budget. This fraction demonstrates that filaments represent a significant component of the galaxy’s gaseous content, with important implications for the mass distribution and cycling of circumgalactic gas.

\subsubsection{Position-velocity diagrams along filaments}

\begin{figure*}[bthp!]
\includegraphics[width=1.0\textwidth]{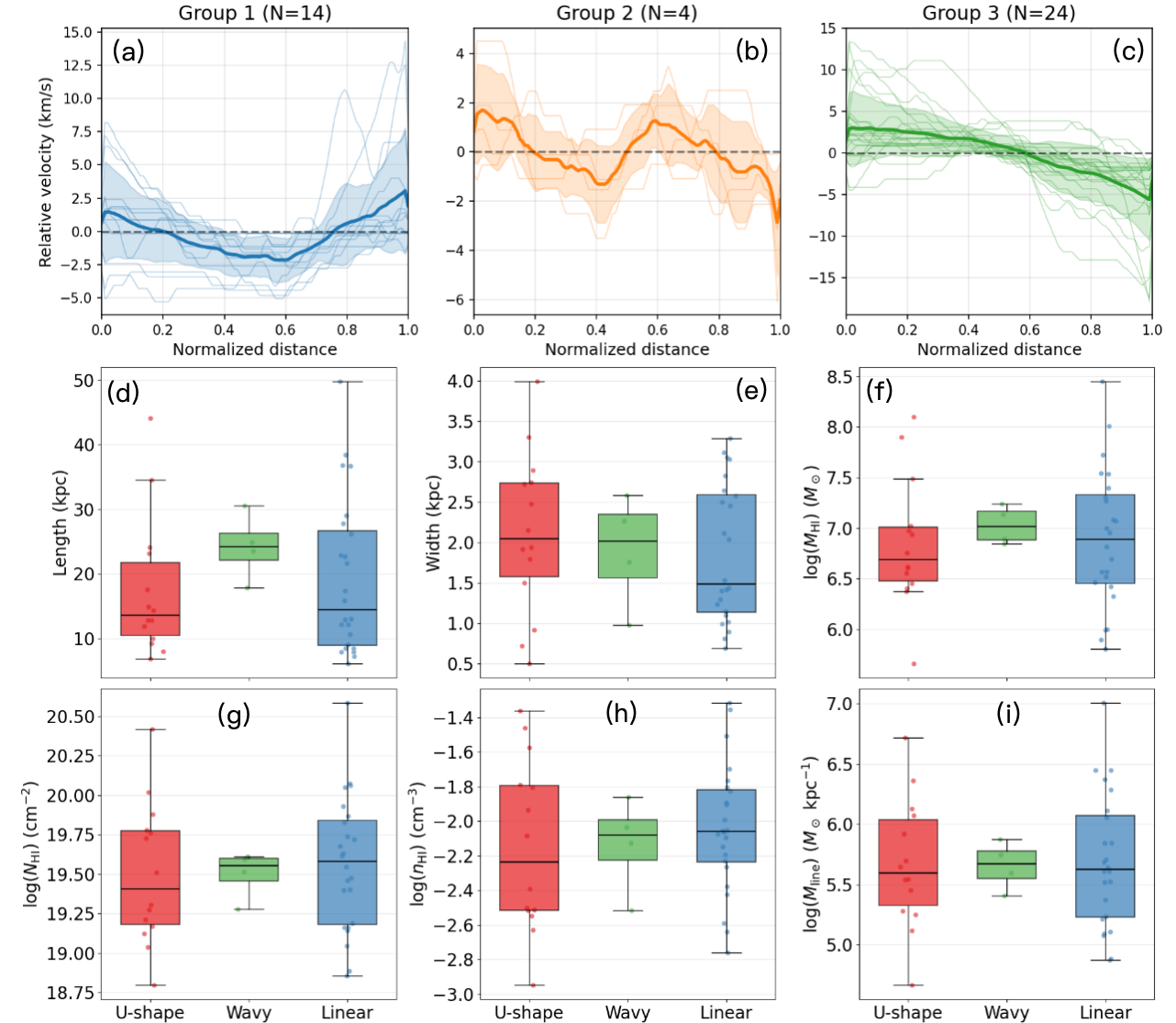}
\caption{Panel (a)-(c): Normalized velocity profiles of the three filament classes derived from DTW clustering. Light curves represent the PV profiles of individual filaments (normalized and de-meaned to remove absolute velocity shifts), the thick solid curve denotes the class-averaged profile, and the shaded region indicates $\pm 1\sigma$ standard deviation. The three classes correspond to the U-shaped profile (outflow), linear profile (inflow), and wavy profile (turbulence), respectively. The horizontal axis represents the normalized arc length along the filament skeleton, and the vertical axis represents the relative line-of-sight velocity. Panel (d)-(i): Distributions of physical parameters for the three filament classes shown as box plots. Panels (d) to (i) sequentially show: filament length, filament width, $\log M_{\rm HI}$, $\log N_{\rm HI}$, $\log n_{\rm HI}$, and $\log M_{\rm line}$. In each panel, the red box corresponds to U-shaped outflow filaments ($N=14$), the green box corresponds to wavy turbulent filaments ($N=4$), and the blue box corresponds to linear inflow filaments ($N=24$). The box represents the IQR, the horizontal line inside the box marks the median, the whiskers extend to the data range excluding outliers, and the scatter points represent the raw data of individual filaments. The horizontal axis labels indicate the kinematic morphology and its physical interpretation, and the vertical axis represents the corresponding physical quantity.
}
\label{fig:PV_cluster}
\end{figure*}

Fig.~\ref{fig:density_profile}(b) shows the position-velocity (PV) diagram of an examplar filament, which describes velocity variation along its filament skeleton. A PV diagram is generated for each filament. Using dynamic time warping (DTW) to compute pairwise distances between all filaments (comparing curve shapes rather than absolute positions), we perform clustering analysis to classify filaments into three kinematic subgroups: U-shaped profile filaments ($N=14$), wavy-shaped profile filaments ($N=4$), and linear-shaped profile filaments ($N=24$). Figures~\ref{fig:PV_cluster}(a)-(c) present typical PV curves from each cluster.

Different PV morphologies may reflect distinct kinematic features. U-shaped profiles may represent global collapse of the filament, with the filament converging toward the center and contracting radially inward \citep{2022MNRAS.514.6038Z,2024A&A...689A..74A}. Linear PV profiles exhibit a monotonic gradient without velocity reversal, possibly representing unidirectional steady inflow toward the galactic potential well, or alternatively, outgoing tidal tails dragged by companion galaxies. Wavy PV profiles display irregular random undulations along the filament, possibly indicating small-scale turbulence and localized random motions. Given the small number of wavy filaments ($N=4$), their statistical results should be interpreted with caution.

To investigate what physical factors influence the different kinematic morphologies of the filaments, we perform statistical analyses (median $\pm$ interquartile range) and Kruskal-Wallis tests on various physical parameters across the three classes. Results are shown in Figures~\ref{fig:PV_cluster}(d)-(i). Global Kruskal-Wallis tests reveal no significant differences ($p > 0.05$) in intrinsic physical parameters (length, width, $\log M_{\rm HI}$, $\log N_{\rm HI}$, $\log n_{\rm HI}$, $\log M_{\rm line}$) among the three groups, indicating they belong to the same statistical distribution overall. In other word, the velocity profile morphology of HI filaments is not dominated by the intrinsic physical properties (geometric scale, $M_{\rm HI}$, $N_{\rm HI}$, $n_{\rm HI}$, and $M_{\rm line}$). 

The three kinematic classes of filamentary structures show prominent spatial segregation in the circumgalactic and intergalactic medium (CGM/IGM). Filaments with linear position-velocity (PV) profiles preferentially trace galactic tidal tails, implying that their formation and morphology are primarily governed by large-scale tidal interactions. In contrast, U-shaped and wavy filaments predominantly reside in the diffuse outer CGM/IGM, outside the tidally regulated vicinity of the galactic disk where direct tidal perturbations are negligible. We nevertheless identify a small set of exceptional sources: filaments 6 and 47 are categorized as U-shaped filaments yet reside within the northern tidal tail, deviating from the general spatial trend of their population. However, this overall spatial dichotomy demonstrates that galactic tidal fields cannot universally dominate the origin and dynamical evolution of all CGM/IGM filamentary structures.

\begin{figure*}[bthp!]
\includegraphics[width=1.0\textwidth]{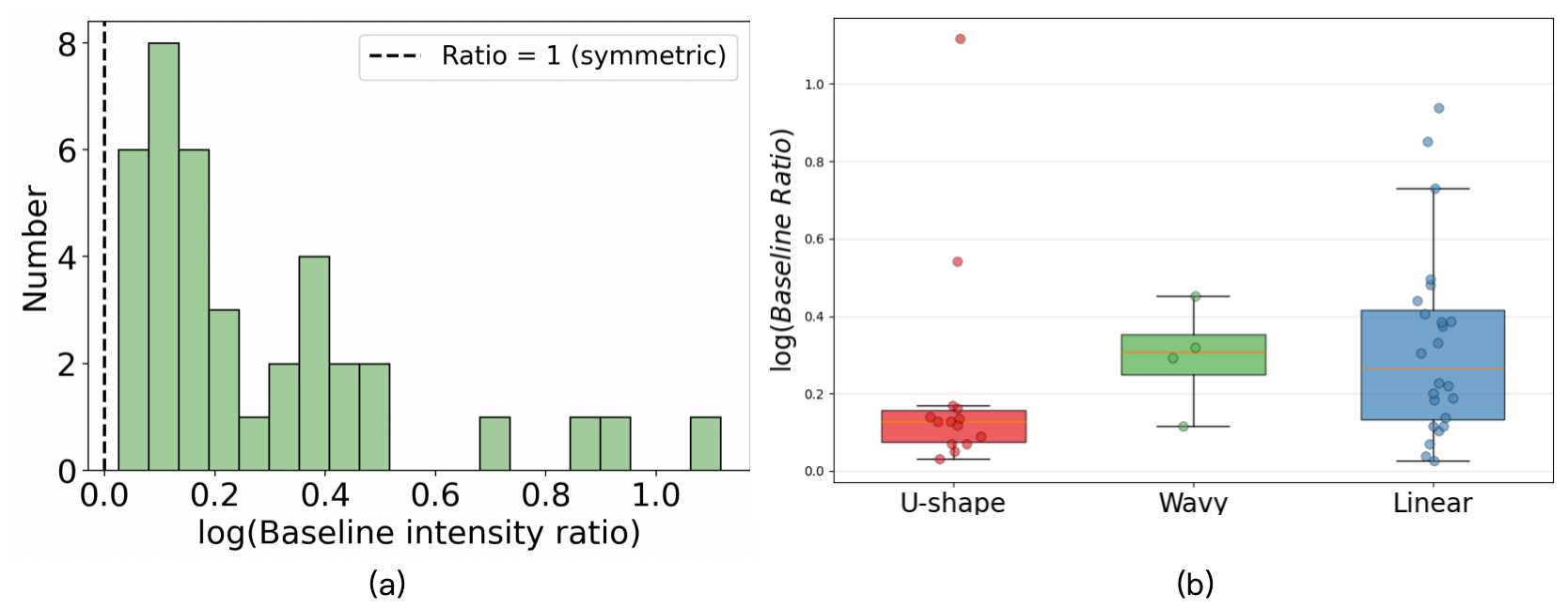}
\caption{Panel (a): Histogram of the density-profile baseline asymmetry. The horizontal axis shows the logarithmic baseline ratio $\log(\text{Baseline Ratio})$, and the vertical axis denotes the number of filaments. The black dashed line marks the symmetric baseline position at a ratio of unity ($\log=0$). Panel (b): Box plot of density-profile baseline asymmetry for the three filament populations. The horizontal axis represents PV morphological types, including U-shaped outflow filaments ($N=14$), wavy turbulent filaments ($N=4$), and linear inflow filaments ($N=24$); the vertical axis corresponds to the logarithmic baseline ratio $\log(\text{Baseline Ratio})$. Each box indicates the IQR, the horizontal line inside each box denotes the median, and the scattered points show the raw values of individual filaments.}
\label{fig:baseline}
\end{figure*}

\section{Discussions}

The above analysis demonstrates that filamentary structures are pervasive throughout the circumgalactic medium (CGM) and intragalactic medium (IGM), yet the physical origin of these gaseous filaments remains poorly constrained. Filaments exhibiting linear position-velocity (PV) profiles are most likely shaped by tidal interactions. In contrast, filaments with U-shaped PV profiles share striking similarities with star-forming filaments in molecular clouds. On parsec scales, U-shaped or V-shaped PV signatures are ubiquitous in star-forming filaments and are generally interpreted as products of gravitational collapse along filamentary structures \citep{2022MNRAS.514.6038Z,2024A&A...689A..74A}. However, the CGM and IGM are distinguished by extremely low gas densities and tenuous ambient conditions, which differ drastically from the dense molecular cloud environments of star-forming regions. Under such diffuse circumstances, gravitational forces are insufficient to drive efficient global collapse of gas structures. Consequently, the convergent kinematic signatures traced by U-shaped CGM/IGM filaments cannot be gravitationally induced; instead, they likely originate from non-gravitational mechanisms, primarily local ram pressure compression and turbulent inertial motions. This physical scenario is further supported by the spatial concentration of U-shaped filaments across the CGM/IGM, including regions where tidal interactions are present.

During baseline subtraction of the filament density profiles, we identify systematic left-right baseline asymmetry in a subset of filamentary sources. This asymmetry encodes meaningful physical information, hinting for unilateral compressive motions of filamentary gas. We quantify this asymmetry via a left-to-right baseline ratio, as visualized in Figure~\ref{fig:baseline}(a). We further characterize baseline asymmetry across our PV-based filament classification scheme in Figure~\ref{fig:baseline}(b). The magnitude of baseline asymmetry in density-profile varies distinctly across the three filament populations. The median logarithmic baseline asymmetry values are $0.128$ for U-shaped filaments, $0.306$ for wavy filaments, and $0.266$ for linear filaments. The global Kruskal-Wallis test yields a marginally significant statistical result of $p=0.1018$, and pairwise comparisons confirm a significant difference between linear and U-shaped filaments ($p=0.0442$). These systematic statistical differences further imply diverse formation mechanisms for IGM/CGM filament populations. After excluding two outliers (sources 6 and 47), the remaining U-shaped filaments exhibit consistently low baseline asymmetry, indicating the absence of a dominant unidirectional drag or compression. Considering that the overall environment of NGC 4631 is tidally driven \citep{Combes1978A&A....65...47C}, we interpret these U-shaped filaments as tracers of localized hydrodynamic complexity within this tidal framework. Specifically, they may have experienced stripping, colission, compression, and turbulent cascading that deviate from bulk unidirectional flow. These filaments form localized high-density regions, potentially further enhanced with thermal instability and rapid cooling in hot halo.

In this work, we report the first direct observational detection of the long-hypothesized kpc-scale filamentary network within the CGM/IGM. This component has long represented a critical missing link in the cosmic baryon cycle. It seems that elongated filamentary structures pervade cosmic structures across vastly different spatial scales: Mpc-scale cosmic web, kpc-scale circumgalactic/intergalactic medium (CGM/IGM), and parsec-scale interstellar medium (ISM). Despite arising from fundamentally distinct physical environments, these filamentary structures share striking morphological similarities, implying that filament assembly across all scales could be regulated by some universal physical mechanisms---most notably anisotropic compression. Future more dedicated investigations of filamentary structures threading the CGM and IGM, empowered by state-of-the-art facilities including the SKA, are poised to yield unprecedented observational constraints that deepen our theoretical picture of the cosmic baryon ecosystem, especially the baryonic mass exchange between galaxies and their ambient circumgalactic and intergalactic environments.

\begin{acknowledgments}
This work has been supported by the National Science and Technology Major Project of China (No. 2024ZD1100601) and National Key R\&D Program of China (No. 2022YFA1603100). Z.C. acknowledges support from the National Natural Science Foundation of China (NSFC), through grants No. 12403028, the Basic Research Program of Shanxi Provence (202403021222272). T.L. acknowledges support from the National Natural Science Foundation of China (NSFC), through grants No. 12073061 and No. 12122307, the Tianchi Talent Program of Xin-jiang Uygur Autonomous Region, and the international partnership program of the Chinese Academy of Sciences, through grant No. 114231KYSB20200009.

\end{acknowledgments}

\appendix
\setcounter{figure}{0}
\renewcommand{\thefigure}{A\arabic{figure}}
\section{Filaments in the Disks of NGC 4631 and NGC 4656}
In the main text, we focus on the filamentary structures detected in the CGM and IGM regions surrounding NGC 4631. As shown in Figure~\ref{fig:skeleton}, two prominent filaments are also identified within the galactic disks of NGC 4631 and NGC 4656. These disk filaments possess distinctly different physical properties from the CGM/IGM filaments, and are therefore discussed separately in this appendix. Spatially, the two filaments coincide well with the active star-forming regions inside the two galaxies, implying a close physical connection between filamentary gas and ongoing star formation activity. The filament in NGC 4631 extends along the major axis of the galactic disk, and the filament in NGC 4656 follows a similar orientation. Both filaments are embedded within the high-column-density H{\,\small I} galactic disk and do not belong to the diffuse CGM/IGM component.

The neutral hydrogen in the galactic disk is not a single uniform component, but consists of two components with distinct physical properties. This conclusion is derived from our fitting analysis of the transverse density profiles of the galactic disk. Specifically, when extracting transverse density profiles for the disk region, the disk profiles exhibit a characteristic profile with a central sharp peak and broad wings on both sides, which is a typical signature of a two-component distribution. Accordingly, we adopt a double Gaussian model for fitting. The derived widths and corresponding uncertainties of the two Gaussian components are listed in Table~\ref{tab:filament_params}. The plots of the two Gaussian fittings are presented in Figures A1 and A2.

Notably, the disk filaments of NGC 4631 are not completely spatially isolated from the CGM/IGM filaments. As can be seen in Figure~\ref{fig:skeleton}, some CGM/IGM filaments appear to connect with the ends of the disk filaments, suggesting a potential physical linkage between them.

\begin{figure*}[bthp!]
\centering
\includegraphics[width=\textwidth]{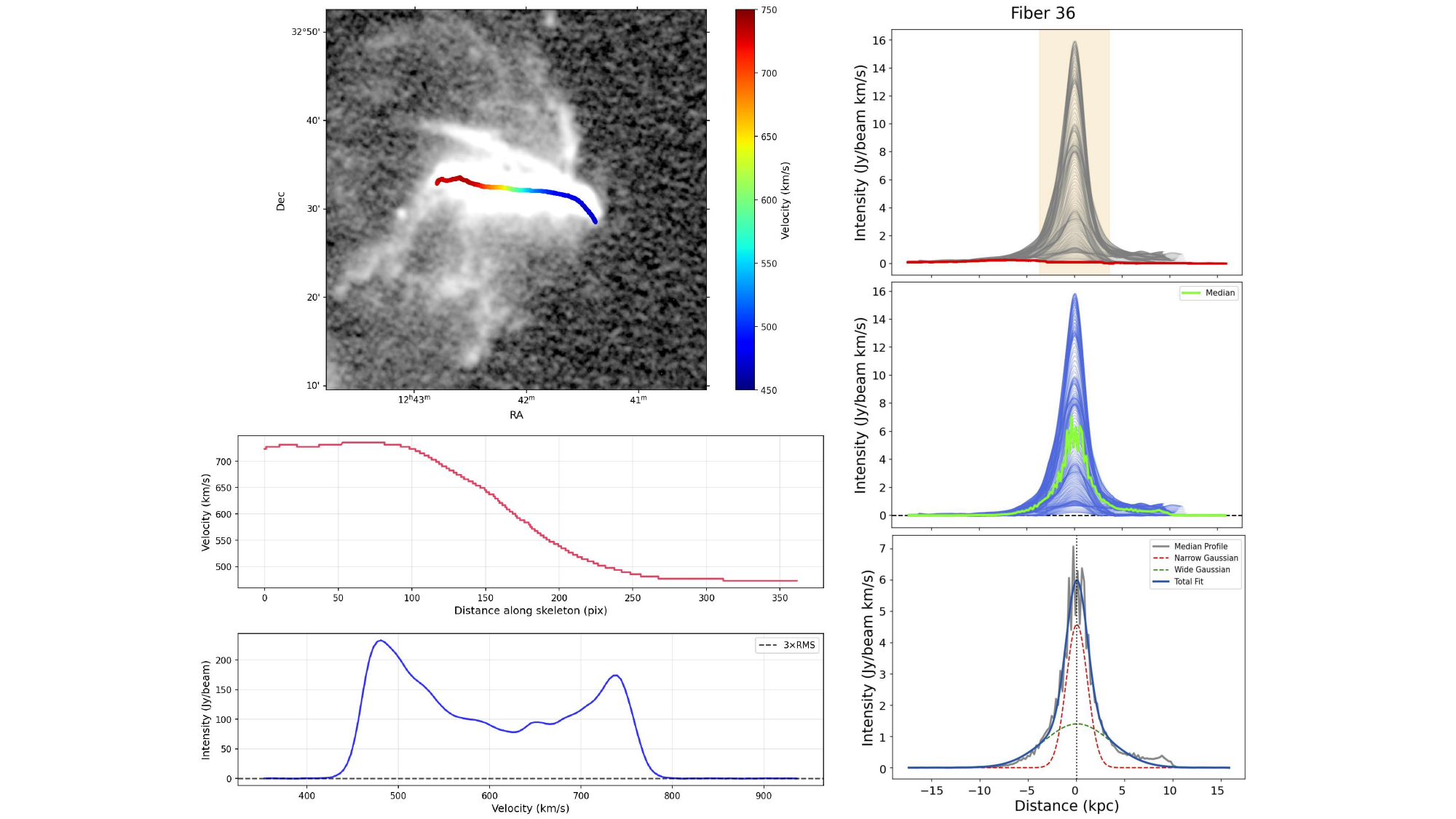}
\caption{Keep the legend consistent with Figure~\ref{fig:density_profile}.}
\label{fig:app_a1}
\end{figure*}

\begin{figure*}[bthp!]
\centering
\includegraphics[width=\textwidth]{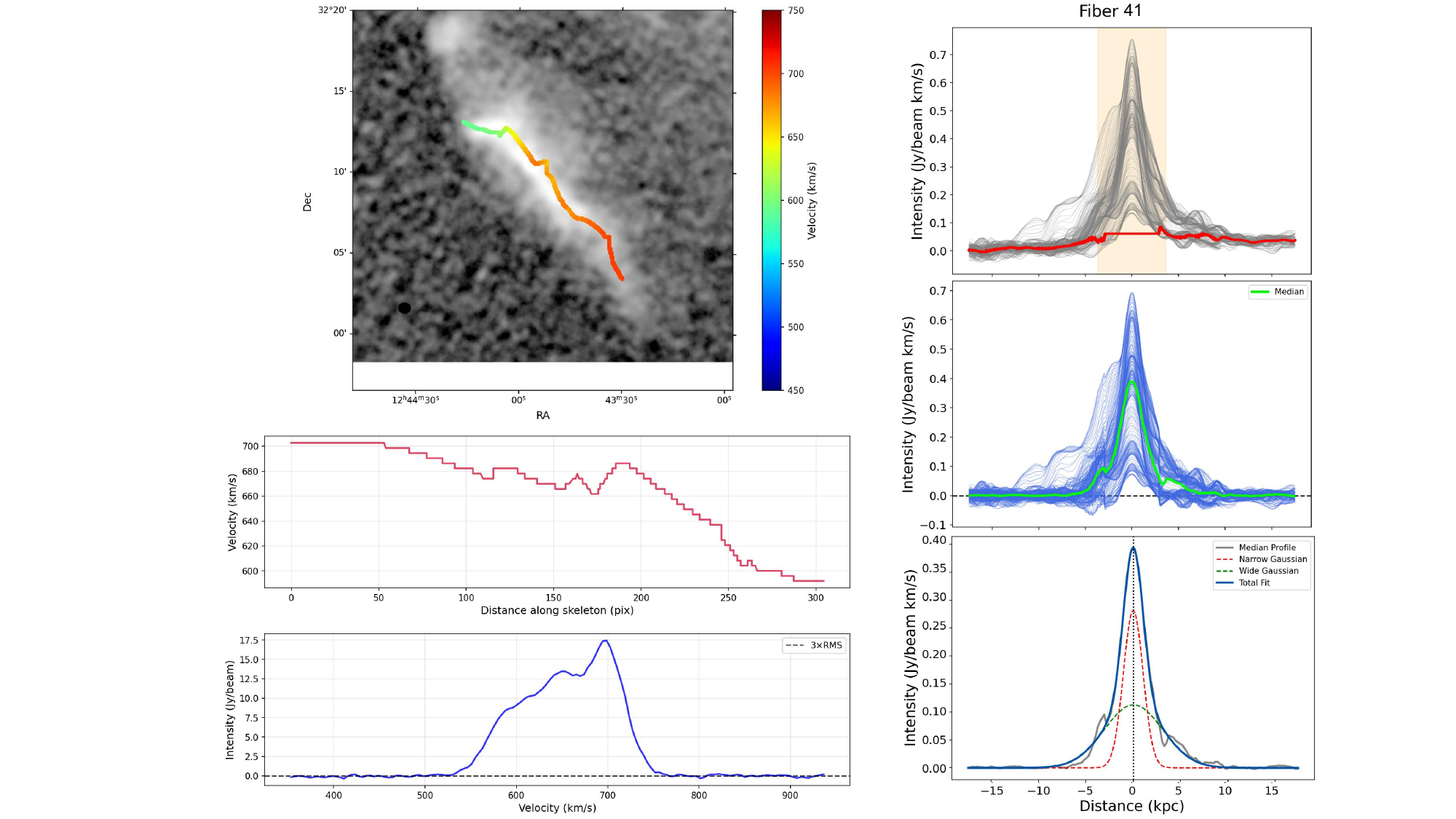}
\caption{Keep the legend consistent with Figure~\ref{fig:density_profile}.}
\label{fig:app_a2}
\end{figure*}

\section{Physical parameters of NGC 4631 filaments}

We assume that the filament has a uniform structure and compute the average column density using the Gaussian fitting results of the median integrated intensity profile. The abscissa $x$ of this profile is the spatial distance along the filament width direction (in units of kpc), and the ordinate is the velocity-integrated flux density at each spatial position:

\[
I(x) = \int S(x, v) \, dv \quad [\text{Jy/beam} \cdot \text{km/s}]
\]

We perform a spatial integration over the entire profile:

\[
\mathcal{I} = \int_{0}^{W} I(x) \, dx \quad [\text{Jy/beam} \cdot \text{km/s} \cdot \text{kpc}]
\]

where $W$ is the projected width of the filament (in units of kpc). We then convert the total integrated flux to column density. The main-beam brightness temperature $T_{\text{mb}}$ of the H{\,\small I} line is related to the flux density $S$ by:

\[
T_{\text{mb}} = \frac{\lambda^2}{2 k_{\text{B}} \Omega_{\text{beam}}} S \times 10^{-26}
\]

where $\lambda = 0.2110611405 \, \text{m}$ is the wavelength of the H{\,\small I} line, $k_{\text{B}} = 1.380649 \times 10^{-23} \, \text{J} \cdot \text{K}^{-1}$ is the Boltzmann constant, and $\Omega_{\text{beam}}$ is the beam solid angle. For a Gaussian beam, the solid angle is computed from the half-power beam widths:

\[
\Omega_{\text{beam}} = 1.1331 \, \theta_{\text{maj}} \, \theta_{\text{min}}
\]

where $\theta_{\text{maj}}$ and $\theta_{\text{min}}$ are the major and minor axes of the beam, respectively (in units of radians). The integrated main-beam temperature is:

\[
\int T_{\text{mb}} \, dv = \frac{\lambda^2}{2 k_{\text{B}} \Omega_{\text{beam}}} \cdot \mathcal{I} \times 10^{-26}
\]

The relationship between H{\,\small I} column density and integrated main-beam temperature is:

\[
N_{\text{HI, total}} = C \int T_{\text{mb}} \, dv
\]

where $C = 1.8224 \times 10^{18} \, \text{cm}^{-2} / (\text{K} \cdot \text{km} \cdot \text{s}^{-1})$ is a constant. Combining the above equations, the total column density of the median profile is:

\[
N_{\text{HI, total}} = C \cdot \frac{\lambda^2}{2 k_{\text{B}} \Omega_{\text{beam}}} \cdot \mathcal{I} \times 10^{-26}
\]

Since $\mathcal{I}$ includes the integration along the width direction, $N_{\text{HI, total}}$ has dimensions of column density multiplied by length ($\text{cm}^{-2} \cdot \text{kpc}$). Therefore, the average column density of the filament is:

\[
\langle N_{\text{HI}} \rangle = \frac{N_{\text{HI, total}}}{W} \quad [\text{cm}^{-2}]
\]

The total H{\,\small I} mass of the filament is calculated by multiplying the average column density by the projected area:

\[
M_{\text{HI}} = \frac{m_{\text{H}}}{M_{\odot}} \cdot \langle N_{\text{HI}} \rangle \cdot W \cdot L \cdot (1 \, \text{kpc})^2
\]

where $m_{\text{H}} = 1.6737236 \times 10^{-27} \, \text{kg}$ is the hydrogen atom mass, $M_{\odot} = 1.988475415 \times 10^{30} \, \text{kg}$ is the solar mass, $1 \, \text{kpc} = 3.085677581 \times 10^{21} \, \text{cm}$, $W$ and $L$ are the projected width and length of the filament (in units of kpc).

We assume that the filament has a cylindrical geometry with its major axis lying in the plane of the sky, and that the observational line of sight is perpendicular to the major axis. Under this geometric assumption, the physical path length of the line of sight through the filament equals the projected width of the filament:

\[
\Delta s = W \quad [\text{kpc}]
\]

where $W$ in units of kpc. Converting the path length to centimeters:

\[
\Delta s_{\text{cm}} = W \times 3.085677581 \times 10^{21} \quad [\text{cm}]
\]

The average H{\,\small I} volume density inside the filament is then obtained by dividing the average column density by the line-of-sight path length:

\[
n_{\text{HI}} = \frac{\langle N_{\text{HI}} \rangle}{\Delta s_{\text{cm}}} \quad [\text{cm}^{-3}] .
\]

\begin{table*}
\centering
\caption{Physical parameters of NGC 4631 filaments}
\label{tab:filament_params}
\resizebox{\textwidth}{!}{%
\begin{tabular}{ccccccccc}
\hline
Filament ID & Width (kpc) & Length (kpc) & $N_\mathrm{HI}$ (cm$^{-2}$) & $n_\mathrm{HI}$ (cm$^{-3}$) & $M_\mathrm{HI}$ ($M_\odot$) & $M_\mathrm{line}$ ($M_\odot$ kpc$^{-1}$) & Class \\
\hline
1 & 2.113 $\pm$ 0.119 & 29.049 & 4.12e+19 & 8.05e-03 & 2.03e+07 & 6.98e+05 & L \\
2 & 1.293 $\pm$ 0.112 & 10.678 & 7.16e+18 & 2.29e-03 & 7.93e+05 & 7.42e+04 & L \\
3 & 2.892 $\pm$ 0.512 & 23.128 & 5.77e+19 & 8.23e-03 & 3.09e+07 & 1.34e+06 & U \\
4 & 1.097 $\pm$ 0.104 & 26.207 & 1.45e+19 & 5.45e-03 & 3.34e+06 & 1.27e+05 & L \\
5 & 1.501 $\pm$ 0.100 & 14.361 & 1.48e+19 & 4.07e-03 & 2.56e+06 & 1.78e+05 & U \\
6* & 0.718 $\pm$ 1.571 & 6.813 & 6.04e+19 & 3.47e-02 & 2.37e+06 & 3.47e+05 & U \\
7 & 0.992 $\pm$ 0.985 & 17.367 & 1.54e+19 & 6.42e-03 & 2.13e+06 & 1.23e+05 & L \\
8 & 1.797 $\pm$ 0.163 & 14.904 & 1.33e+19 & 3.04e-03 & 2.84e+06 & 1.91e+05 & U \\
9 & 2.646 $\pm$ 0.185 & 15.837 & 1.11e+19 & 1.73e-03 & 3.73e+06 & 2.36e+05 & L \\
10 & 2.717 $\pm$ 0.070 & 9.188 & 2.03e+19 & 3.08e-03 & 4.05e+06 & 4.41e+05 & U \\
11 & 2.829 $\pm$ 0.137 & 12.906 & 8.55e+19 & 1.25e-02 & 2.50e+07 & 1.94e+06 & L \\
12 & 3.053 $\pm$ 0.155 & 9.074 & 5.28e+19 & 7.13e-03 & 1.17e+07 & 1.29e+06 & L \\
13 & 0.809 $\pm$ 0.191 & 6.114 & 2.51e+19 & 1.28e-02 & 9.95e+05 & 1.63e+05 & L \\
14 & 1.016 $\pm$ 0.518 & 8.460 & 1.45e+19 & 5.91e-03 & 1.00e+06 & 1.18e+05 & L \\
15 & 1.761 $\pm$ 0.179 & 24.816 & 3.94e+19 & 9.24e-03 & 1.38e+07 & 5.57e+05 & W \\
16 & 3.301 $\pm$ 0.354 & 17.578 & 1.88e+19 & 2.35e-03 & 8.74e+06 & 4.97e+05 & U \\
17 & 0.689 $\pm$ 0.218 & 12.171 & 7.37e+19 & 4.42e-02 & 4.95e+06 & 4.07e+05 & L \\
18 & 1.530 $\pm$ 0.243 & 21.628 & 1.39e+19 & 3.76e-03 & 3.70e+06 & 1.71e+05 & L \\
19 & 2.584 $\pm$ 0.188 & 17.829 & 1.91e+19 & 3.04e-03 & 7.03e+06 & 3.95e+05 & W \\
20* & 1.939 $\pm$ 0.326 & 7.978 & 7.60e+19 & 1.62e-02 & 9.43e+06 & 1.18e+06 & U \\
21 & 1.437 $\pm$ 0.161 & 7.230 & 3.53e+19 & 1.01e-02 & 2.94e+06 & 4.06e+05 & L \\
22 & 1.236 $\pm$ 0.185 & 8.428 & 7.72e+18 & 2.58e-03 & 6.44e+05 & 7.64e+04 & L \\
23 & 1.147 $\pm$ 0.254 & 22.890 & 4.77e+19 & 1.72e-02 & 1.00e+07 & 4.39e+05 & L \\
24* & 3.989 $\pm$ 0.946 & 11.883 & 1.09e+19 & 1.13e-03 & 4.13e+06 & 3.48e+05 & U \\
25 & 2.573 $\pm$ 0.120 & 7.935 & 5.52e+19 & 8.86e-03 & 9.04e+06 & 1.14e+06 & L \\
26 & 3.110 $\pm$ 0.063 & 12.200 & 1.13e+20 & 1.50e-02 & 3.43e+07 & 2.81e+06 & L \\
27 & 0.503 $\pm$ 0.079 & 44.056 & 3.24e+19 & 2.66e-02 & 5.75e+06 & 1.31e+05 & U \\
28 & 1.400 $\pm$ 0.095 & 7.926 & 2.99e+19 & 8.80e-03 & 2.65e+06 & 3.35e+05 & L \\
29 & 0.917 $\pm$ 0.190 & 9.979 & 6.28e+18 & 2.83e-03 & 4.61e+05 & 4.62e+04 & U \\
30* & 2.267 $\pm$ 0.177 & 23.488 & 4.09e+19 & 7.44e-03 & 1.75e+07 & 7.43e+05 & W \\
31 & 2.500 $\pm$ 0.122 & 13.033 & 2.53e+19 & 4.18e-03 & 6.61e+06 & 5.07e+05 & L \\
32 & 2.153 $\pm$ 0.164 & 12.787 & 1.64e+19 & 3.14e-03 & 3.62e+06 & 2.83e+05 & U \\
33 & 1.916 $\pm$ 0.125 & 12.797 & 5.37e+19 & 1.16e-02 & 1.06e+07 & 8.25e+05 & U \\
34 & 3.029 $\pm$ 0.101 & 36.719 & 1.15e+20 & 1.56e-02 & 1.02e+08 & 2.79e+06 & L \\
35* & 2.288 $\pm$ 0.896 & 29.917 & 1.46e+20 & 2.63e-02 & 7.99e+07 & 2.67e+06 & - \\
36** & 2.099 $\pm$ 0.139 & 44.162 & 3.79e+21 & 7.44e-01 & 2.81e+09 & 6.37e+07 & - \\
     & 8.008 $\pm$ 0.806 & 44.162 & 9.53e+20 & 4.91e-02 & 2.70e+09 & 6.12e+07 & - \\
37 & 3.285 $\pm$ 0.109 & 27.841 & 3.83e+20 & 4.81e-02 & 2.81e+08 & 1.01e+07 & L \\
38 & 1.418 $\pm$ 0.077 & 36.824 & 2.90e+19 & 8.44e-03 & 1.21e+07 & 3.30e+05 & L \\
39 & 2.456 $\pm$ 0.101 & 22.654 & 1.19e+20 & 2.00e-02 & 5.30e+07 & 2.34e+06 & L \\
40* & 3.053 $\pm$ 0.155 & 42.642 & 5.28e+19 & 7.13e-03 & 5.51e+07 & 1.29e+06 & - \\
41** & 1.912 $\pm$ 0.058 & 35.624 & 2.40e+20 & 5.18e-02 & 1.31e+08 & 3.68e+06 & - \\
     & 7.116 $\pm$ 0.250 & 35.624 & 7.69e+19 & 4.46e-03 & 1.56e+08 & 4.39e+06 & - \\
42 & 0.976 $\pm$ 0.149 & 30.582 & 3.27e+19 & 1.38e-02 & 7.82e+06 & 2.56e+05 & W \\
43 & 2.741 $\pm$ 0.105 & 34.541 & 1.04e+20 & 1.57e-02 & 7.90e+07 & 2.29e+06 & U \\
44 & 0.896 $\pm$ 1.295 & 38.423 & 6.78e+19 & 3.12e-02 & 1.87e+07 & 4.87e+05 & L \\
45* & 2.333 $\pm$ 0.135 & 34.306 & 6.84e+19 & 1.21e-02 & 4.39e+07 & 1.28e+06 & - \\
46 & 2.036 $\pm$ 0.109 & 49.762 & 4.28e+19 & 8.67e-03 & 3.47e+07 & 6.98e+05 & L \\
47 & 2.477 $\pm$ 0.065 & 24.086 & 2.62e+20 & 4.36e-02 & 1.25e+08 & 5.20e+06 & U \\
\hline
\end{tabular}%
}
\vspace{4pt}
\noindent
\textit{Note.} — Class denotes the kinematic morphology of each filament based on its PV diagram: \textbf{U} = U-shaped profile, \textbf{W} = wavy-shaped profile, \textbf{L} = linear-shaped profile. Filaments with half-profiles due to disk contamination are marked with an asterisk(*), and the two disk filaments are denoted by double asterisk(**).

\end{table*}


\bibliography{sample701}{}

@ARTICLE{2025NatAs...9.1366L,
       author = {{Liu}, Xunchuan and {Liu}, Tie and {Li}, Pak-Shing and {Mai}, Xiaofeng and {Henkel}, Christian and {Goldsmith}, Paul F. and {Qin}, Sheng-Li and {Gong}, Yan and {Lu}, Xing and {Xu}, Fengwei and {Luo}, Qiuyi and {Liu}, Hong-Li and {Zhang}, Tianwei and {Cheng}, Yu and {Di}, Yihuan and {Wu}, Yuefang and {Gu}, Qilao and {Tang}, Ningyu and {Yang}, Aiyuan and {Shen}, Zhiqiang},
        title = "{A network of velocity-coherent filaments formed by supersonic turbulence in a very-high-velocity H I cloud}",
      journal = {Nature Astronomy},
         year = 2025,
        month = sep,
       volume = {9},
        pages = {1366-1374},
          doi = {10.1038/s41550-025-02605-8},
archivePrefix = {arXiv},
       eprint = {2502.10897},
 primaryClass = {astro-ph.GA},
       adsurl = {https://ui.adsabs.harvard.edu/abs/2025NatAs...9.1366L}
}

@ARTICLE{2010A&A...518L.102A,
       author = {{Andr{\'e}}, Ph. and {Men'shchikov}, A. and {Bontemps}, S. and {K{\"o}nyves}, V. and {Motte}, F. and {Schneider}, N. and {Didelon}, P. and {Minier}, V. and {Saraceno}, P. and {Ward-Thompson}, D. and {di Francesco}, J. and {White}, G. and {Molinari}, S. and {Testi}, L. and {Abergel}, A. and {Griffin}, M. and {Henning}, Th. and {Royer}, P. and {Mer{\'\i}n}, B. and {Vavrek}, R. and {Attard}, M. and {Arzoumanian}, D. and {Wilson}, C.~D. and {Ade}, P. and {Aussel}, H. and {Baluteau}, J.-P. and {Benedettini}, M. and {Bernard}, J.-Ph. and {Blommaert}, J.~A.~D.~L. and {Cambr{\'e}sy}, L. and {Cox}, P. and {di Giorgio}, A. and {Hargrave}, P. and {Hennemann}, M. and {Huang}, M. and {Kirk}, J. and {Krause}, O. and {Launhardt}, R. and {Leeks}, S. and {Le Pennec}, J. and {Li}, J.~Z. and {Martin}, P.~G. and {Maury}, A. and {Olofsson}, G. and {Omont}, A. and {Peretto}, N. and {Pezzuto}, S. and {Prusti}, T. and {Roussel}, H. and {Russeil}, D. and {Sauvage}, M. and {Sibthorpe}, B. and {Sicilia-Aguilar}, A. and {Spinoglio}, L. and {Waelkens}, C. and {Woodcraft}, A. and {Zavagno}, A.},
        title = "{From filamentary clouds to prestellar cores to the stellar IMF: Initial highlights from the Herschel Gould Belt Survey}",
      journal = {\aap},
         year = 2010,
        month = jul,
       volume = {518},
          eid = {L102},
        pages = {L102},
          doi = {10.1051/0004-6361/201014666},
archivePrefix = {arXiv},
       eprint = {1005.2618},
 primaryClass = {astro-ph.GA},
       adsurl = {https://ui.adsabs.harvard.edu/abs/2010A&A...518L.102A}
}

@INPROCEEDINGS{2014prpl.conf...27A,
       author = {{Andr{\'e}}, P. and {Di Francesco}, J. and {Ward-Thompson}, D. and {Inutsuka}, S.-I. and {Pudritz}, R.~E. and {Pineda}, J.~E.},
        title = "{From Filamentary Networks to Dense Cores in Molecular Clouds: Toward a New Paradigm for Star Formation}",
    booktitle = {Protostars and Planets VI},
         year = 2014,
       editor = {{Beuther}, Henrik and {Klessen}, Ralf S. and {Dullemond}, Cornelis P. and {Henning}, Thomas},
        month = jan,
        pages = {27-51},
          doi = {10.2458/azu_uapress_9780816531240-ch002},
archivePrefix = {arXiv},
       eprint = {1312.6232},
 primaryClass = {astro-ph.GA},
       adsurl = {https://ui.adsabs.harvard.edu/abs/2014prpl.conf...27A}
}

@ARTICLE{2026arXiv260211617Z,
       author = {{Zhang}, Yan-Kun and {Liu}, Tie and {Jiao}, Wenyu and {Li}, Pak-Shing and {Zeng}, Jia and {Zhang}, Chao and {Garc{\'\i}a}, Pablo and {Juvela}, Mika and {Garay}, Guido and {Stutz}, Amelia M. and {Dib}, Sami and {Meng}, Dezhao and {Feng}, Jian-Cheng and {Yang}, Dongting and {Xu}, Fengwei and {Tej}, Anandmayee and {V{\'a}zquez-Semadeni}, Enrique and {G{\'o}mez}, Gilberto C. and {Zhang}, Yong and {Tang}, Xindi and {Goldsmith}, Paul F. and {Kim}, Kee-Tae and {Chibueze}, James O. and {Ren}, Zhiyuan and {Sanhueza}, Patricio and {Yang}, Aiyuan and {Hwang}, Jihye and {Li}, Shanghuo and {Baug}, Tapas and {Gupta}, Shivani and {Das}, Swagat R. and {Wu}, Gang and {Zhou}, Jianjun and {Lee}, Chang Won and {Dewangan}, Lokesh and {Gorai}, Prasanta and {Lyu}, Tianning and {Zhu}, Lei},
        title = "{The ALMA-QUARKS Survey: Discovery of Dusty Fibrils inside Massive Star-forming Clumps}",
      journal = {arXiv e-prints},
         year = 2026,
        month = feb,
          eid = {arXiv:2602.11617},
        pages = {arXiv:2602.11617},
          doi = {10.48550/arXiv.2602.11617},
archivePrefix = {arXiv},
       eprint = {2602.11617},
 primaryClass = {astro-ph.GA},
       adsurl = {https://ui.adsabs.harvard.edu/abs/2026arXiv260211617Z}
}

@ARTICLE{2026arXiv260404501Z,
       author = {{Zhang}, Chao and {Liu}, Tie and {Juvela}, Mika and {Padoan}, Paolo and {Liu}, Hong-Li and {Li}, Di and {Garay}, Guido and {Evans}, Neal J. and {Xu}, Fengwei and {Goldsmith}, Paul F. and {Zhang}, Qizhou and {Kim}, Kee-Tae and {Zhang}, Yankun and {Ren}, Zhiyuan and {Zhao}, Mengke},
        title = "{Random gas motions inside sub-parsec scale supercritical filaments}",
      journal = {arXiv e-prints},
         year = 2026,
        month = apr,
          eid = {arXiv:2604.04501},
        pages = {arXiv:2604.04501},
          doi = {10.48550/arXiv.2604.04501},
archivePrefix = {arXiv},
       eprint = {2604.04501},
 primaryClass = {astro-ph.GA},
       adsurl = {https://ui.adsabs.harvard.edu/abs/2026arXiv260404501Z}
}

@ARTICLE{2022MNRAS.514.6038Z,
       author = {{Zhou}, Jian-Wen and {Liu}, Tie and {Evans}, Neal J. and {Garay}, Guido and {Goldsmith}, Paul F. and {G{\'o}mez}, Gilberto C. and {V{\'a}zquez-Semadeni}, Enrique and {Liu}, Hong-Li and {Stutz}, Amelia M. and {Wang}, Ke and {Juvela}, Mika and {He}, Jinhua and {Li}, Di and {Bronfman}, Leonardo and {Liu}, Xunchuan and {Xu}, Feng-Wei and {Tej}, Anandmayee and {Dewangan}, L.~K. and {Li}, Shanghuo and {Zhang}, Siju and {Zhang}, Chao and {Ren}, Zhiyuan and {Tatematsu}, Ken'ichi and {Shing Li}, Pak and {Won Lee}, Chang and {Baug}, Tapas and {Qin}, Sheng-Li and {Wu}, Yuefang and {Peng}, Yaping and {Zhang}, Yong and {Liu}, Rong and {Luo}, Qiu-Yi and {Ge}, Jixing and {Saha}, Anindya and {Chakali}, Eswaraiah and {Zhang}, Qizhou and {Kim}, Kee-Tae and {Ristorcelli}, Isabelle and {Shen}, Zhi-Qiang and {Li}, Jin-Zeng},
        title = "{ATOMS: ALMA Three-millimeter Observations of Massive Star-forming regions - XI. From inflow to infall in hub-filament systems}",
      journal = {\mnras},
         year = 2022,
        month = aug,
       volume = {514},
       number = {4},
        pages = {6038-6052},
          doi = {10.1093/mnras/stac1735},
archivePrefix = {arXiv},
       eprint = {2206.08505},
 primaryClass = {astro-ph.GA},
       adsurl = {https://ui.adsabs.harvard.edu/abs/2022MNRAS.514.6038Z}
}

@ARTICLE{2024A&A...689A..74A,
       author = {{{\'A}lvarez-Guti{\'e}rrez}, R.~H. and {Stutz}, A.~M. and {Sandoval-Garrido}, N. and {Louvet}, F. and {Motte}, F. and {Galv{\'a}n-Madrid}, R. and {Cunningham}, N. and {Sanhueza}, P. and {Bonfand}, M. and {Bontemps}, S. and {Gusdorf}, A. and {Ginsburg}, A. and {Csengeri}, T. and {Reyes}, S.~D. and {Salinas}, J. and {Baug}, T. and {Bronfman}, L. and {Busquet}, G. and {D{\'\i}az-Gonz{\'a}lez}, D.~J. and {Fernandez-Lopez}, M. and {Guzm{\'a}n}, A. and {Koley}, A. and {Liu}, H.-L. and {Olguin}, F.~A. and {Valeille-Manet}, M. and {Wyrowski}, F.},
        title = "{ALMA-IMF: XIII. N$_{2}$H$^{+}$ kinematic analysis of the intermediate protocluster G353.41}",
      journal = {\aap},
         year = 2024,
        month = sep,
       volume = {689},
          eid = {A74},
        pages = {A74},
          doi = {10.1051/0004-6361/202450321},
archivePrefix = {arXiv},
       eprint = {2404.07363},
 primaryClass = {astro-ph.GA},
       adsurl = {https://ui.adsabs.harvard.edu/abs/2024A&A...689A..74A}
}

@ARTICLE{Wang2023ApJ...944..102W,
       author = {{Wang}, Jing and {Yang}, Dong and {Oh}, S.-H. and {Staveley-Smith}, Lister and {Wang}, Jie and {Wang}, Q. Daniel and {Hess}, Kelley M. and {Ho}, Luis C. and {Hou}, Ligang and {Jing}, Yingjie and {Kamphuis}, Peter and {Li}, Fujia and {Lin}, Xuchen and {Liu}, Ziming and {Shao}, Li and {Wang}, Shun and {Zhu}, Ming},
        title = "{FEASTS: IGM Cooling Triggered by Tidal Interactions through the Diffuse H I Phase around NGC 4631}",
      journal = {\apj},
         year = 2023,
        month = feb,
       volume = {944},
       number = {1},
          eid = {102},
        pages = {102},
          doi = {10.3847/1538-4357/acafe8},
archivePrefix = {arXiv},
       eprint = {2301.00937},
 primaryClass = {astro-ph.GA},
       adsurl = {https://ui.adsabs.harvard.edu/abs/2023ApJ...944..102W}
}

@ARTICLE{Lin2025ApJ...995...12L,
       author = {{Lin}, Zeren and {Martin}, D. Christopher and {Matuszewski}, Mateusz and {Neill}, James D. and {Miles}, Drew M. and {Picouet}, Vincent and {Prusinski}, Nikolaus Z. and {Hoadley}, Keri and {McGurk}, Rosalie},
        title = "{Kinematically Complex Circumgalactic Gas Around a Low-mass Galaxy: Filamentary Inflow and Counterrotation in J0910b}",
      journal = {\apj},
         year = 2025,
        month = dec,
       volume = {995},
       number = {1},
          eid = {12},
        pages = {12},
          doi = {10.3847/1538-4357/ae10b2},
       adsurl = {https://ui.adsabs.harvard.edu/abs/2025ApJ...995...12L}
}

@ARTICLE{Melendez2015ApJ...804...46M,
       author = {{Mel{\'e}ndez}, M. and {Veilleux}, S. and {Martin}, C. and {Engelbracht}, C. and {Bland-Hawthorn}, J. and {Cecil}, G. and {Heitsch}, F. and {McCormick}, A. and {M{\"u}ller}, T. and {Rupke}, D. and {Teng}, S.~H.},
        title = "{Exploring the Dust Content of Galactic Winds with Herschel. I. NGC 4631}",
      journal = {\apj},
         year = 2015,
        month = may,
       volume = {804},
       number = {1},
          eid = {46},
        pages = {46},
          doi = {10.1088/0004-637X/804/1/46},
archivePrefix = {arXiv},
       eprint = {1502.07785},
 primaryClass = {astro-ph.GA},
       adsurl = {https://ui.adsabs.harvard.edu/abs/2015ApJ...804...46M}
}

@INPROCEEDINGS{Hacar2023ASPC..534..153H,
       author = {{Hacar}, A. and {Clark}, S.~E. and {Heitsch}, F. and {Kainulainen}, J. and {Panopoulou}, G.~V. and {Seifried}, D. and {Smith}, R.},
        title = "{Initial Conditions for Star Formation: a Physical Description of the Filamentary ISM}",
    booktitle = {Protostars and Planets VII},
         year = 2023,
       editor = {{Inutsuka}, S. and {Aikawa}, Y. and {Muto}, T. and {Tomida}, K. and {Tamura}, M.},
       series = {Astronomical Society of the Pacific Conference Series},
       volume = {534},
        month = jul,
        pages = {153},
          doi = {10.48550/arXiv.2203.09562},
archivePrefix = {arXiv},
       eprint = {2203.09562},
 primaryClass = {astro-ph.GA},
       adsurl = {https://ui.adsabs.harvard.edu/abs/2023ASPC..534..153H}
}

@article{Keres10.1111/j.1365-2966.2005.09451.x,
    author = {Kereš, Dušan and Katz, Neal and Weinberg, David H. and Davé, Romeel},
    title = {How do galaxies get their gas?},
    journal = {Monthly Notices of the Royal Astronomical Society},
    volume = {363},
    number = {1},
    pages = {2-28},
    year = {2005},
    month = {10},
    issn = {0035-8711},
    doi = {10.1111/j.1365-2966.2005.09451.x},
    url = {https://doi.org/10.1111/j.1365-2966.2005.09451.x},
    eprint = {https://academic.oup.com/mnras/article-pdf/363/1/2/4126209/363-1-2.pdf},
}

@ARTICLE{Putman2012ARA&A..50..491P,
       author = {{Putman}, M.~E. and {Peek}, J.~E.~G. and {Joung}, M.~R.},
        title = "{Gaseous Galaxy Halos}",
      journal = {\araa},
         year = 2012,
        month = sep,
       volume = {50},
        pages = {491-529},
          doi = {10.1146/annurev-astro-081811-125612},
archivePrefix = {arXiv},
       eprint = {1207.4837},
 primaryClass = {astro-ph.GA},
       adsurl = {https://ui.adsabs.harvard.edu/abs/2012ARA&A..50..491P}
}

@ARTICLE{Tumlinson2011ApJ...733..111T,
       author = {{Tumlinson}, J. and {Werk}, J.~K. and {Thom}, C. and {Meiring}, J.~D. and {Prochaska}, J.~X. and {Tripp}, T.~M. and {O'Meara}, J.~M. and {Okrochkov}, M. and {Sembach}, K.~R.},
        title = "{Multiphase Gas in Galaxy Halos: The O VI Lyman-limit System toward J1009+0713}",
      journal = {\apj},
         year = 2011,
        month = jun,
       volume = {733},
       number = {2},
          eid = {111},
        pages = {111},
          doi = {10.1088/0004-637X/733/2/111},
archivePrefix = {arXiv},
       eprint = {1103.5252},
 primaryClass = {astro-ph.CO},
       adsurl = {https://ui.adsabs.harvard.edu/abs/2011ApJ...733..111T}
}

@ARTICLE{Saintonge2017ApJS..233...22S,
       author = {{Saintonge}, Am{\'e}lie and {Catinella}, Barbara and {Tacconi}, Linda J. and {Kauffmann}, Guinevere and {Genzel}, Reinhard and {Cortese}, Luca and {Dav{\'e}}, Romeel and {Fletcher}, Thomas J. and {Graci{\'a}-Carpio}, Javier and {Kramer}, Carsten and {Heckman}, Timothy M. and {Janowiecki}, Steven and {Lutz}, Katharina and {Rosario}, David and {Schiminovich}, David and {Schuster}, Karl and {Wang}, Jing and {Wuyts}, Stijn and {Borthakur}, Sanchayeeta and {Lamperti}, Isabella and {Roberts-Borsani}, Guido W.},
        title = "{xCOLD GASS: The Complete IRAM 30 m Legacy Survey of Molecular Gas for Galaxy Evolution Studies}",
      journal = {\apjs},
         year = 2017,
        month = dec,
       volume = {233},
       number = {2},
          eid = {22},
        pages = {22},
          doi = {10.3847/1538-4365/aa97e0},
archivePrefix = {arXiv},
       eprint = {1710.02157},
 primaryClass = {astro-ph.GA},
       adsurl = {https://ui.adsabs.harvard.edu/abs/2017ApJS..233...22S}
}

@ARTICLE{Borthakur2016ApJ...833..259B,
       author = {{Borthakur}, Sanchayeeta and {Heckman}, Timothy and {Tumlinson}, Jason and {Bordoloi}, Rongmon and {Kauffmann}, Guinevere and {Catinella}, Barbara and {Schiminovich}, David and {Dav{\'e}}, Romeel and {Moran}, Sean M. and {Saintonge}, Amelie},
        title = "{The Properties of the Circumgalactic Medium in Red and Blue Galaxies: Results from the COS-GASS+COS-Halos Surveys}",
      journal = {\apj},
         year = 2016,
        month = dec,
       volume = {833},
       number = {2},
          eid = {259},
        pages = {259},
          doi = {10.3847/1538-4357/833/2/259},
archivePrefix = {arXiv},
       eprint = {1609.06308},
 primaryClass = {astro-ph.GA},
       adsurl = {https://ui.adsabs.harvard.edu/abs/2016ApJ...833..259B}
}

@ARTICLE{Lan2018ApJ...866...36L,
       author = {{Lan}, Ting-Wen and {Mo}, Houjun},
        title = "{The Circumgalactic Medium of eBOSS Emission Line Galaxies: Signatures of Galactic Outflows in Gas Distribution and Kinematics}",
      journal = {\apj},
         year = 2018,
        month = oct,
       volume = {866},
       number = {1},
          eid = {36},
        pages = {36},
          doi = {10.3847/1538-4357/aadc08},
archivePrefix = {arXiv},
       eprint = {1806.05786},
 primaryClass = {astro-ph.GA},
       adsurl = {https://ui.adsabs.harvard.edu/abs/2018ApJ...866...36L}
}

@ARTICLE{Blumenthal2018MNRAS.479.3952B,
       author = {{Blumenthal}, Kelly A. and {Barnes}, Joshua E.},
        title = "{Go with the Flow: Understanding inflow mechanisms in galaxy collisions}",
      journal = {\mnras},
         year = 2018,
        month = sep,
       volume = {479},
       number = {3},
        pages = {3952-3965},
          doi = {10.1093/mnras/sty1605},
archivePrefix = {arXiv},
       eprint = {1806.05132},
 primaryClass = {astro-ph.GA},
       adsurl = {https://ui.adsabs.harvard.edu/abs/2018MNRAS.479.3952B}
}

@ARTICLE{Wang2020ApJ...890...63W,
       author = {{Wang}, Jing and {Catinella}, Barbara and {Saintonge}, Am{\'e}lie and {Pan}, Zhizheng and {Serra}, Paolo and {Shao}, Li},
        title = "{xGASS: H I Fueling of Star Formation in Disk-dominated Galaxies}",
      journal = {\apj},
         year = 2020,
        month = feb,
       volume = {890},
       number = {1},
          eid = {63},
        pages = {63},
          doi = {10.3847/1538-4357/ab68dd},
archivePrefix = {arXiv},
       eprint = {2001.01970},
 primaryClass = {astro-ph.GA},
       adsurl = {https://ui.adsabs.harvard.edu/abs/2020ApJ...890...63W}
}

@ARTICLE{Yu2022ApJ...930...85Y,
       author = {{Yu}, Niankun and {Ho}, Luis C. and {Wang}, Jing},
        title = "{Centrally Concentrated H I Distribution Enhances Star Formation in Galaxies}",
      journal = {\apj},
         year = 2022,
        month = may,
       volume = {930},
       number = {1},
          eid = {85},
        pages = {85},
          doi = {10.3847/1538-4357/ac5f07},
archivePrefix = {arXiv},
       eprint = {2203.13405},
 primaryClass = {astro-ph.GA},
       adsurl = {https://ui.adsabs.harvard.edu/abs/2022ApJ...930...85Y}
}

@ARTICLE{Meng2026ApJ..1002..210M,
       author = {{Meng}, Yuxi and {Wang}, Jie and {Jing}, Yingjie and {Chen}, Hongxiang and {Liu}, Zerui},
        title = "{Detection of H I Filaments: Pair Stacking versus Filament Stacking}",
      journal = {\apj},
         year = 2026,
        month = may,
       volume = {1002},
       number = {2},
          eid = {210},
        pages = {210},
          doi = {10.3847/1538-4357/ae61a5},
archivePrefix = {arXiv},
       eprint = {2509.22180},
 primaryClass = {astro-ph.CO},
       adsurl = {https://ui.adsabs.harvard.edu/abs/2026ApJ..1002..210M}
}

@article{CHENdoi:10.1142/S2010194512006459,
author = {CHEN, XUELEI},
title = {THE TIANLAI PROJECT: A 21CM COSMOLOGY EXPERIMENT},
journal = {International Journal of Modern Physics: Conference Series},
volume = {12},
number = {},
pages = {256-263},
year = {2012},
doi = {10.1142/S2010194512006459},

URL = { 
    
        https://doi.org/10.1142/S2010194512006459
    
    

},
eprint = { 
    
        https://doi.org/10.1142/S2010194512006459
    
    

}
}

@article{Li_2023,
doi = {10.3847/1538-4357/ace896},
url = {https://doi.org/10.3847/1538-4357/ace896},
year = {2023},
month = {aug},
publisher = {The American Astronomical Society},
volume = {954},
number = {2},
pages = {139},
author = {Li, Yichao and Wang, Yougang and Deng, Furen and Yang, Wenxiu and Hu, Wenkai and Liu, Diyang and Zhao, Xinyang and Zuo, Shifan and Shu, Shuanghao and Li, Jixia and Timbie, Peter and Ansari, Réza and Perdereau, Olivier and Stebbins, Albert and Wolz, Laura and Wu, Fengquan and Zhang, Xin and Chen, Xuelei},
title = {FAST Drift Scan Survey for Hi Intensity Mapping: I. Preliminary Data Analysis},
journal = {The Astrophysical Journal}
}

@ARTICLE{Zucker2015ApJ...815...23Z,
       author = {{Zucker}, Catherine and {Battersby}, Cara and {Goodman}, Alyssa},
        title = "{The Skeleton of the Milky Way}",
      journal = {\apj},
         year = 2015,
        month = dec,
       volume = {815},
       number = {1},
          eid = {23},
        pages = {23},
          doi = {10.1088/0004-637X/815/1/23},
archivePrefix = {arXiv},
       eprint = {1506.08807},
 primaryClass = {astro-ph.GA},
       adsurl = {https://ui.adsabs.harvard.edu/abs/2015ApJ...815...23Z}
}

@ARTICLE{Ragan2014A&A...568A..73R,
       author = {{Ragan}, S.~E. and {Henning}, Th. and {Tackenberg}, J. and {Beuther}, H. and {Johnston}, K.~G. and {Kainulainen}, J. and {Linz}, H.},
        title = "{Giant molecular filaments in the Milky Way}",
      journal = {\aap},
         year = 2014,
        month = aug,
       volume = {568},
          eid = {A73},
        pages = {A73},
          doi = {10.1051/0004-6361/201423401},
archivePrefix = {arXiv},
       eprint = {1403.1450},
 primaryClass = {astro-ph.GA},
       adsurl = {https://ui.adsabs.harvard.edu/abs/2014A&A...568A..73R}
}

@ARTICLE{Zhang2019A&A...622A..52Z,
       author = {{Zhang}, M. and {Kainulainen}, J. and {Mattern}, M. and {Fang}, M. and {Henning}, Th.},
        title = "{Star-forming content of the giant molecular filaments in the Milky Way}",
      journal = {\aap},
         year = 2019,
        month = feb,
       volume = {622},
          eid = {A52},
        pages = {A52},
          doi = {10.1051/0004-6361/201732400},
archivePrefix = {arXiv},
       eprint = {1811.02197},
 primaryClass = {astro-ph.GA},
       adsurl = {https://ui.adsabs.harvard.edu/abs/2019A&A...622A..52Z}
}

@ARTICLE{Colombo2021A&A...655L...2C,
       author = {{Colombo}, D. and {K{\"o}nig}, C. and {Urquhart}, J.~S. and {Wyrowski}, F. and {Mattern}, M. and {Menten}, K.~M. and {Lee}, M.-Y. and {Brand}, J. and {Wienen}, M. and {Mazumdar}, P. and {Schuller}, F. and {Leurini}, S.},
        title = "{OGHReS: Large-scale filaments in the outer Galaxy}",
      journal = {\aap},
         year = 2021,
        month = nov,
       volume = {655},
          eid = {L2},
        pages = {L2},
          doi = {10.1051/0004-6361/202142182},
archivePrefix = {arXiv},
       eprint = {2111.02768},
 primaryClass = {astro-ph.GA},
       adsurl = {https://ui.adsabs.harvard.edu/abs/2021A&A...655L...2C}
}

@ARTICLE{Veena2021ApJ...921L..42V,
       author = {{Veena}, V.~S. and {Schilke}, P. and {S{\'a}nchez-Monge}, {\'A}. and {Sormani}, M.~C. and {Klessen}, R.~S. and {Schuller}, F. and {Colombo}, D. and {Csengeri}, T. and {Mattern}, M. and {Urquhart}, J.~S.},
        title = "{A Kiloparsec-scale Molecular Wave in the Inner Galaxy: Feather of the Milky Way?}",
      journal = {\apjl},
         year = 2021,
        month = nov,
       volume = {921},
       number = {2},
          eid = {L42},
        pages = {L42},
          doi = {10.3847/2041-8213/ac341f},
archivePrefix = {arXiv},
       eprint = {2110.13938},
 primaryClass = {astro-ph.GA},
       adsurl = {https://ui.adsabs.harvard.edu/abs/2021ApJ...921L..42V}
}

@ARTICLE{Jiang2019SCPMA..6259502J,
       author = {{Jiang}, Peng and {Yue}, YouLing and {Gan}, HengQian and {Yao}, Rui and {Li}, Hui and {Pan}, GaoFeng and {Sun}, JingHai and {Yu}, DongJun and {Liu}, HongFei and {Tang}, NingYu and {Qian}, Lei and {Lu}, JiGuang and {Yan}, Jun and {Peng}, Bo and {Zhang}, ShuXin and {Wang}, QiMing and {Li}, Qi and {Li}, Di and {FAST Collaboration}},
        title = "{Commissioning progress of the FAST}",
      journal = {Science China Physics, Mechanics, and Astronomy},
         year = 2019,
        month = may,
       volume = {62},
       number = {5},
          eid = {959502},
        pages = {959502},
          doi = {10.1007/s11433-018-9376-1},
archivePrefix = {arXiv},
       eprint = {1903.06324},
 primaryClass = {astro-ph.IM},
       adsurl = {https://ui.adsabs.harvard.edu/abs/2019SCPMA..6259502J}
}

@ARTICLE{Mohapatra2022MNRAS.514.3139M,
       author = {{Mohapatra}, Rajsekhar and {Federrath}, Christoph and {Sharma}, Prateek},
        title = "{Multiphase turbulence in galactic haloes: effect of the driving}",
      journal = {\mnras},
         year = 2022,
        month = aug,
       volume = {514},
       number = {3},
        pages = {3139-3159},
          doi = {10.1093/mnras/stac1610},
archivePrefix = {arXiv},
       eprint = {2206.03602},
 primaryClass = {astro-ph.GA},
       adsurl = {https://ui.adsabs.harvard.edu/abs/2022MNRAS.514.3139M}
}

@ARTICLE{OSullivan2014ApJ...793...74O,
       author = {{O'Sullivan}, E. and {Vrtilek}, J.~M. and {David}, L.~P. and {Giacintucci}, S. and {Zezas}, A. and {Ponman}, T.~J. and {Mamon}, G.~A. and {Nulsen}, P. and {Raychaudhury}, S.},
        title = "{Deep Chandra Observations of HCG 16. II. The Development of the Intra-group Medium in a Spiral-rich Group}",
      journal = {\apj},
         year = 2014,
        month = oct,
       volume = {793},
       number = {2},
          eid = {74},
        pages = {74},
          doi = {10.1088/0004-637X/793/2/74},
archivePrefix = {arXiv},
       eprint = {1407.7546},
 primaryClass = {astro-ph.CO},
       adsurl = {https://ui.adsabs.harvard.edu/abs/2014ApJ...793...74O}
}

@ARTICLE{Jones2019,
       author = {{Jones}, M.~G. and {Verdes-Montenegro}, L. and {Damas-Segovia}, A. and {Borthakur}, S. and {Yun}, M. and {del Olmo}, A. and {Perea}, J. and {Rom{\'a}n}, J. and {Luna}, S. and {Lopez Gutierrez}, D. and {Williams}, B. and {Vogt}, F.~P.~A. and {Garrido}, J. and {Sanchez}, S. and {Cannon}, J. and {Ram{\'\i}rez-Moreta}, P.},
        title = "{Evolution of compact groups from intermediate to final stages. A case study of the H I content of HCG 16}",
      journal = {\aap},
         year = 2019,
        month = dec,
       volume = {632},
          eid = {A78},
        pages = {A78},
          doi = {10.1051/0004-6361/201936349},
archivePrefix = {arXiv},
       eprint = {1910.03420},
 primaryClass = {astro-ph.GA},
       adsurl = {https://ui.adsabs.harvard.edu/abs/2019A&A...632A..78J}
}

@ARTICLE{Heald2011A&A...526A.118H,
       author = {{Heald}, G. and {J{\'o}zsa}, G. and {Serra}, P. and {Zschaechner}, L. and {Rand}, R. and {Fraternali}, F. and {Oosterloo}, T. and {Walterbos}, R. and {J{\"u}tte}, E. and {Gentile}, G.},
        title = "{The Westerbork Hydrogen Accretion in LOcal GAlaxieS (HALOGAS) survey. I. Survey description and pilot observations}",
      journal = {\aap},
         year = 2011,
        month = feb,
       volume = {526},
          eid = {A118},
        pages = {A118},
          doi = {10.1051/0004-6361/201015938},
archivePrefix = {arXiv},
       eprint = {1012.0816},
 primaryClass = {astro-ph.CO},
       adsurl = {https://ui.adsabs.harvard.edu/abs/2011A&A...526A.118H}
}

@ARTICLE{Combes1978A&A....65...47C,
       author = {{Combes}, F.},
        title = "{A model of tidal interactions within the NGC 4631 group of galaxies.}",
      journal = {\aap},
         year = 1978,
        month = apr,
       volume = {65},
        pages = {47-55},
       adsurl = {https://ui.adsabs.harvard.edu/abs/1978A&A....65...47C}
}
\bibliographystyle{aasjournalv7}



\end{document}